\documentclass[11pt]{article}
\usepackage[english]{babel}
\usepackage[a4paper,top=2cm,bottom=2cm,left=2.5cm,right=2.5cm,marginparwidth=1.75cm]{geometry}
\usepackage{setspace}
\usepackage{enumerate}
\usepackage[english]{babel}

\usepackage{amsmath}
\usepackage{graphicx}
\usepackage{natbib}
\usepackage{amssymb}

\usepackage[most]{tcolorbox} 

\definecolor{myblue}{RGB}{173,216,230} 

\usepackage{multicol}

\usepackage[colorlinks=true, allcolors=blue]{hyperref}

\usepackage{authblk}

\makeatletter

\renewcommand\AB@affilsepx{, }

\def\myinlineaffils{%
  {\small
   \renewcommand{\and}{, }
   \AB@affillist}
}

\makeatletter

\renewcommand\AB@affilsepx{, }

\def\myinlineaffils{%
  {\small
   \renewcommand{\and}{, }
   \AB@affillist}
} 
\renewcommand{\maketitle}{%
  \begin{center}
    {\LARGE\@title\par}
    \vskip 1em
    {\large
      \setlength{\parskip}{0.3em}
      \setlength{\parindent}{0pt}
      \@author
      \par}%
    \vskip 1em
    \vskip 1em
  \end{center}
}
\makeatother

\title{Why the Northern Hemisphere Needs a 30–40 m Telescope and the Science at Stake: Cosmology and High-z Universe}

\author[1]{Pablo G. P\'erez-Gonz\'alez}
\author[2]{Roberto Maiolino}
\author[3]{Pascal A. Oesch}
\author[4]{Alvio Renzini}
\author[5]{Tommaso Treu}
\author[6]{Cristina Ramos Almeida}
\author[7]{Sandra Faber}
\author[1]{Luis Colina}
\author[8]{Almudena Alonso-Herrero}
\author[1]{Santiago Arribas}
\author[9]{Guillermo Barro}
\author[6]{Helmut Dannerbauer}
\author[10]{Mark Dickinson}
\author[11]{Mauro Giavalisco}
\author[6]{Marc Huertas-Company}
\author[12]{G\"oran \"Ostlin}
\author[13]{Giulia Rodighiero}
\author[14]{Patricia S\'anchez-Bl\'azquez}
\author[9]{Elisa Toloba}

\affil[1]{Centro de Astrobiolog\'{\i}a (CAB), CSIC-INTA, Ctra. de Ajalvir km 4, Torrej\'on de Ardoz, E-28850, Madrid, Spain}
\affil[2]{Kavli Institute for Cosmology, University of Cambridge, Madingley Road, Cambridge, CB3 0HA, UK}
\affil[3]{Department of Astronomy, University of Geneva, Chemin Pegasi 51, CH-1290 Versoix, Switzerland}
\affil[4]{INAF - Osservatorio Astronomico di Padova, Vicolo dell'Osservatorio 5, I-35122 Padova, Italy}
\affil[5]{Department of Physics and Astronomy, University of California, Los Angeles, CA 90095, USA}
\affil[6]{Instituto de Astrofísica de Canarias, Calle Vía L\'actea, s/n, E-38205, La Laguna, Tenerife, Spain}
\affil[7]{UCO/Lick Observatory, Department of Astronomy and Astrophysics, University of California, Santa Cruz, CA 95064, USA}
\affil[8]{Centro de Astrobiología (CAB), CSIC-INTA, Camino Bajo del Castillo s/n, E-28692 Villanueva de la Cañada, Madrid, Spain}
\affil[9]{Department of Physics, University of the Pacific, 3601 Pacific Avenue, Stockton, CA 95211, USA}
\affil[10]{NSF NOIRLab, 950 N.\ Cherry Ave., Tucson, AZ, 85719, USA}
\affil[11]{Astronomy Department, University of Massachusetts Amherst, MA 01003, USA}
\affil[12]{Department of Astronomy, Stockholm University, Oscar Klein Centre, AlbaNova University Centre, 106 91 Stockholm, Sweden}
\affil[13]{Dipartimento di Fisica e Astronomia "G. Galilei", Universit\`a di Padova, Vicolo dell'Osservatorio 3, 35122 Padova, Italy}
\affil[14]{Departamento de Física de la Tierra y Astrofísica \& Insituto de Física de Partículas y Ciencias del Cosmos (IPARCOS), UCM, E-28040, Madrid, Spain}

\begin{document}
\maketitle

\newpage
\begin{tcolorbox}[
    colback=myblue!20!white, 
    colframe=myblue,         
    arc=5mm,                 
    boxsep=2mm,              
    left=2mm, right=2mm, top=2mm, bottom=2mm 
]
Full sky coverage with 30-40 meter-class telescopes is essential to answer fundamental questions in Astrophysics, Cosmology, and Physics, such as the composition of the Universe and the formation of the first stars and supermassive black holes.~An ELT/TMT-like telescope in the Northern Hemisphere is a  fundamental and necessary facility to provide multiplexing of observing power, diversity of instrumentation, rapid response, and statistical power required to address the questions and the problems, current and future, unveiled by full sky observatories such as JWST, Euclid, or Roman space telescopes. The Northern ELT/TMT will expedite the study of unique, extreme, rare, transient, and/or high-energy events which will give the most information about fundamental Physics problems in the era of multi-messenger and time-domain Astronomy. 
\end{tcolorbox}

\vspace{-0.6cm}

\section{Introduction: Cosmology challenges}
\vspace{-0.3cm}

The composition of the Universe and the origin of the first stars, supermassive black holes (SMBHs) and galaxies after the Big Bang, are two of the most fundamental questions of science. This paper argues that the deployment of an extremely large optical/infrared telescope (30-40 meters diameter) in the Northern Hemisphere would transform our ability to answer these questions. 

{\bf The composition of the Universe.} In order to explain astronomical observations,  our cosmological model posits that 95\% of the mass-energy budget of the Universe is composed of so-called Dark Matter and Dark Energy. Dark matter is hypothesized to be made of particle(s) beyond the standard model of Particle Physics that do not interact with light, and are likely massive and non- or weakly interactive (``cold'' dark matter; CDM). So far, CDM has eluded direct or indirect detection in ground based detectors and accelerators, while astronomical observations have set limits on its properties such as abundance, self-interaction and baryonic cross section, and free streaming length \citep{2005PhR...405..279B,2020A&A...641A...6P}. Dark energy is responsible for the accelerated expansion of the Universe during the past few billion years. Recent observations by the \href{https://data.desi.lbl.gov/doc/papers/}{DESI} experiment suggest that Dark Energy may not be as simple as the cosmological constant ($\Lambda$), but it may be a dynamical quantity. The discrepancy between the current expansion rate of the Universe and that predicted in standard cosmology from early Universe probes, known as the ``Hubble Tension'' \citep{divalentino2025}, could require further additions to the cosmological model, such as, for example, an early dark energy phase or new particles. As we will discuss in the next section, a 30-40~m telescope in the Northern Hemisphere would enable breakthroughs in the identification of the dark matter particle(s), the clarification of the nature of dark energy, and the resolution of the Hubble Tension.

{\bf The dawn of galaxies and SMBHs.}
In the dark matter - dark energy ($\Lambda$CDM) cosmological model, galaxies arise hierarchically from primordial fluctuations enhanced by gravity over cosmic time.  However, we still do not know how and when the first stars and galaxies formed. Furthermore, although SMBHs are found ubiquitously at the centers of nearby galaxies, it is not known whether they form from stellar remnants or from direct collapse, and whether they predate or co-evolve with galaxies. A parallel and profound question is whether primordial black holes (PBHs) also exist, what is their mass spectrum and what are their connections with the dark sector and Cosmology. 

The existence of very massive galaxies at early cosmic epochs, assembled in an anti-hierarchical way rather than following the bottom-up evolution of dark matter halos, has been known for more than two decades now \citep{2003ApJ...587L..79F}. Although early results from JWST reporting over-massive galaxies at $z=7-9$, which would have formed more stars than possible based on  the current cosmological paradigm \citep{2023NatAs...7..731B}, have been discarded, other results still challenge our view of the early Universe. These might have implications on the $\Lambda$CDM paradigm, even requiring new Physics such as non-standard dark matter, new particles, modified gravity, varying fundamental constants, and/or dynamical dark energy. Among those challenging results,  we highlight the abundance of galaxies with strong emission in the rest-frame ultraviolet (typically a strong continuum and some relatively faint lines) discovered by JWST in the first 500~Myr of cosmic history, which exceeds expectations of galaxy formation simulations by at least an order of magnitude. The results have been achieved with different datasets, by different teams around the world, and have been corroborated after confirming the high redshift nature of many galaxies with spectroscopy (see references in \citep{2025NatAs...9.1134A}). The current redshift frontier is $z\sim14.5$, i.e., galaxies that existed just 280~Myr after the Big Bang \citep{2025arXiv250511263N}. Photometric candidates now exist up to $z\sim25$ \citep{2025ApJ...991..179P} and even $z\sim30$ \citep{2025arXiv250901664G}, which would place the end of the Dark Ages just 100~Myr after the Big Bang. Possible explanations for this higher than expected abundance of $z\gtrsim10$ galaxies rely on baryon physics, e.g., invoking feedback-free starbursts in primeval galaxies, resulting on higher star formation efficiencies and possible quick metal enrichment and dust production and/or a distinct Initial Mass Function in the early Universe (e.g., \citep{2023MNRAS.523.3201D}).~Other alternatives could have more fundamental implications about the distribution and nature of dark matter \citep{2023JCAP...10..072A,2024ApJ...976...40W,2024MNRAS.530.4868Y,2024MNRAS.527.2835S,2025arXiv250400075S,2025ApJ...988..264M}.


The question about the nature of dark matter links to a second problem with strong cosmological implications and of fundamental origin in Astrophysics. A physical explanation is needed  not only for the high abundance of high redshift galaxies, but also for the ubiquity of active galactic nuclei in the early Universe found by JWST \citep{2024A&A...691A.145M}, with a special attention to a newly discovered type of objects called Little Red Dots (LRDs; \citep{2024ApJ...963..129M}).~The origin of SMBHs, their seeding in the early Universe \citep{2023ApJ...953L..29L}, their influence and, maybe, prominence over star formation at early epochs, and the possible link to PBHs \citep{2025A&A...697A..65M}, most probably have strong implications on the foundations of Physics in the Universe.

\section{The need for a 30-40-meter class telescope in the North}
\vspace{-0.3cm}

The fundamental problems outlined in the previous Section can be addressed through the observations of the distant Universe, but they require higher sensitivity and finer angular resolution than what is enabled by the most powerful telescopes currently operating, including all the top-of-their-class facilities, namely, VLT, Keck, GTC, JWST, and ALMA (see Section 3).

Answering these questions was one of the key scientific drivers of the ESO European Large Telescope (ELT), currently under construction in Chile. We argue that building a comparable (30-40~m diameter) Extremely Large Telescope in the Northern Hemisphere will transform the landscape by enabling access to the entire sky, improving time domain coverage, allowing for independent verification of fundamental results, and providing complementary scientific capabilities, enabling transformative science to solve the open cosmological questions.   

In the nearby Universe (from the Solar System to the Local Group and vicinity), the benefits of full sky coverage are obvious, enabling observations of unique targets.  In the distant universe, the Cosmological Principle ensures that there is no preferred direction and therefore no strictly unique targets.~However, in addition to intrinsically rare targets, there are multiple scientific reasons that make full-sky coverage much more valuable than a single hemisphere.~Those include:

\begin{itemize}
\vspace{-0.3cm}
\item First of all, discovery of new physics requires extraordinary proof. Having two telescopes to carry out independent experiments with different instruments, approaches and techniques will be key to convince the community that indeed new physics might be required.
\vspace{-0.3cm}
\item Unique, extreme, rare, transient, and/or high-energy events are the most promising to provide answers to the fundamental questions presented in the Introduction. Among these special events, we can mention high redshift pair instability supernovae from Population III stars, electromagnetic counterparts or hosts of Gravitational Waves (GWs) sources, implications of high-energy neutrinos, gamma ray bursts at $z>10$ from early stars, and lensed type Ia supernovae. These events probe the seeding and early evolution of stars and SMBHs. In addition, rare and  extreme events include powerful cosmological probes such as standard clocks (gravitational lensing time delays), standard candles (high-z supernovae), and standard sirens (counterparts to GWs).

\vspace{-0.1cm}
\hspace{0.3cm} The list of source types presented above point to the fact that the key in the next decades to probe the composition of the Universe will be multi-messenger time-domain astronomy. Identifying and following-up observations of GWs with observatories such as LIGO, GEO600, VIRGO, KAGRA, the Einstein Telescope, LISA and/or neutrino sources (e.g., those detected by ICECUBE - which are mostly in the North) will require full sky coverage to avoid missing (some of) the most interesting sources among the few events expected every few years. The combination of ELTs in Chile and a 30-40~m class telescope in the North (e.g., in La Palma) will provide a long longitudinal baseline for intra-day time domain astronomy of sources visible from both telescopes. Even though one telescope such as ELT can observe around two thirds of the sky at high enough elevation, at least one observatory in each hemisphere is needed to follow-up intrinsically (rapidly) variable sources for long enough times. We remark that studying the (strong and/or quick) variability of this kind of sources is essential to understand their nature.  
\vspace{-0.3cm}
\item Ongoing and future missions such as JWST, Euclid and Roman, are not restricted to a single hemisphere.~Full sky, fast-response follow-up spectroscopic capability for long-enough periods with a 30-40~m telescope is crucial for fully exploiting these datasets.~In particular, nearly continuous observations of the equatorial belt could be possible with a 30-40~m class telescope in each hemisphere.
\end{itemize}

The uniqueness of some astronomical objects only accessible through an observatory covering the Northern sky is currently demonstrated with sources such as the highest redshift active galactic nuclei, GNz11 \citep{2024Natur.627...59M}, the earliest example of a galaxy where a highly embedded SMBH is clearing its host galaxy's ISM and  quenching star formation, GNz7q \citep{2022Natur.604..261F}, and the prototype of dusty starburst galaxies HDF850.1 \citep{2012Natur.486..233W}, which is supposed to be the main phase in the formation of the largest galaxies known in the local Universe (i.e, ellipticals). This is in part related to the unique fields accessible only from the North, such as the original Hubble Deep Field (extended to the GOODS-N field) or the North Ecliptic Pole field (very important for time-domain observations).

There are other important reasons, at the instrumental and operations level, that also justify an investment in a 30-40~m class telescope in the North: 

\begin{itemize}
\vspace{-0.4cm}
\item Having a second 30-40~m class telescope would also help with diversifying the array of instrumentation, opening different discovery spaces, and doubling, at the very least, the discovery power. This strategy has been shown to be very effective in the four VLTs. In this sense, we remark that ELT is designed to use fully adaptive optics (AO) in a relatively large FoV and optimized for near-infrared observations. Other optic configurations could favor higher Strehl ratios over smaller field of views in the near-IR AO, and be optimized for UV/blue wavelengths in seeing-limited mode.
\vspace{-0.3cm}
\item A pair of telescopes with full sky coverage would enable coordinated Key Programs, with large amounts of time devoted to answering some of the most fundamental and pressing questions, optimizing targets and instrumentation.
\end{itemize}



\vspace{-0.5cm}

\section{Capability Requirements for Solving the Challenge}
\vspace{-0.3cm}

Pushing the frontiers of high redshift science requires breaking barriers in sensitivity and angular resolution, especially for spectroscopic observations. As demonstrated once and again, and most recently by JWST, dramatic improvements in sensitivity lead infallibly to new discoveries.~A few examples of the discoveries that require a 30-40~m class telescope are given below: 

\begin{itemize}
\vspace{-0.3cm}
\item To resolve the sphere of influence of the largest SMBHs at virtually any distance/redshift, thus enabling the most direct and reliable measurements.~Current telescopes (JWST, VLT, GTC, Keck) do not have sufficient angular resolution.
\vspace{-0.3cm}
\item To measure narrow line flux ratios and stellar kinematics of distant gravitational lenses to constrain the nature of dark matter and dark energy \citep[e.g.][]{TSM22,ZNT24}. It would also allow detailed studies of the extreme physical conditions of the early star-forming galaxies with high SNR, e.g., nebular vs. stellar continuum, intervening HI Damped Lyman-$\alpha$ absorption, ionization condition from UV spectral features. Current telescopes do not have sufficient angular resolution and sensitivity.
\vspace{-0.3cm}
\item To resolve the internal structure of galaxies at Cosmic Dawn and Noon to study their internal structure and (dark, baryonic) mass content, which are currently barely resolved with JWST and require a 30-40~m class telescope.
\vspace{-0.3cm}
\item To observe the electromagnetic counterparts to GWs and high-energy neutrinos will require 30$+$-m telescopes for spectroscopy, being too faint and ephemeral for the $<10$-m class. 
\end{itemize}

\vspace{-0.4cm}
More details on the science case and requirements for the study of the composition of the Universe and the dawn of galaxies and SMBHs can be found in the  science case documents for \href{https://www.eso.org/sci/facilities/eelt/science/doc/eelt_sciencecase.pdf}{ELT} and for \href{https://www.tmt.org/download/Document/10/original}{TMT}.

\begin{multicols}{2}
\renewcommand{\bibfont}{\footnotesize}  
\bibliographystyle{plain}
\bibliography{references}

@ARTICLE{2012Natur.486..233W,
       author = {Walter+},
        title = "{}",
      journal = {\nat},
         year = 2012,
       volume = {486},
       number = {7402},
        pages = {233-236},
          doi = {10.1038/nature11073},
archivePrefix = {arXiv},
       eprint = {1206.2641},
 primaryClass = {astro-ph.CO},
       adsurl = {https://ui.adsabs.harvard.edu/abs/2012Natur.486..233W}
}

@ARTICLE{2025ApJ...991..179P,
       author = {P{\'e}rez-Gonz{\'a}lez+},
    title = {},
      journal = {\apj},
         year = 2025,
       volume = {991},
       number = {2},
          eid = {179},
        pages = {179},
          doi = {10.3847/1538-4357/adf8c9},
archivePrefix = {arXiv},
       eprint = {2503.15594},
 primaryClass = {astro-ph.GA},
       adsurl = {https://ui.adsabs.harvard.edu/abs/2025ApJ...991..179P}
}

@ARTICLE{2024ApJ...963..129M,
       author = {Matthee+},
title = {},
      journal = {\apj},
         year = 2024,
       volume = {963},
       number = {2},
          eid = {129},
        pages = {129},
          doi = {10.3847/1538-4357/ad2345},
archivePrefix = {arXiv},
       eprint = {2306.05448},
 primaryClass = {astro-ph.GA},
       adsurl = {https://ui.adsabs.harvard.edu/abs/2024ApJ...963..129M}
}

@ARTICLE{2024A&A...691A.145M,
       author = {Maiolino+},
title = {},
      journal = {\aap},
         year = 2024,
       volume = {691},
          eid = {A145},
        pages = {A145},
          doi = {10.1051/0004-6361/202347640},
archivePrefix = {arXiv},
       eprint = {2308.01230},
 primaryClass = {astro-ph.GA},
       adsurl = {https://ui.adsabs.harvard.edu/abs/2024A&A...691A.145M}
}

@ARTICLE{2023MNRAS.523.3201D,
       author = {Dekel+},
title = {},
      journal = {\mnras},
         year = 2023,
       volume = {523},
       number = {3},
        pages = {3201-3218},
          doi = {10.1093/mnras/stad1557},
archivePrefix = {arXiv},
       eprint = {2303.04827},
 primaryClass = {astro-ph.GA},
       adsurl = {https://ui.adsabs.harvard.edu/abs/2023MNRAS.523.3201D}
}

@ARTICLE{2025arXiv250901664G,
       author = {Gandolfi+},
title = {},
      journal = {arXiv:2509.01664},
         year = 2025,
          eid = {arXiv:2509.01664},
          doi = {10.48550/arXiv.2509.01664},
archivePrefix = {arXiv},
       eprint = {2509.01664},
 primaryClass = {astro-ph.GA},
       adsurl = {https://ui.adsabs.harvard.edu/abs/2025arXiv250901664G}
}

@ARTICLE{2025arXiv250511263N,
       author = {Naidu+},
title = {},
      journal = {arXiv:2505.11263},
         year = 2025,
          eid = {arXiv:2505.11263},
          doi = {10.48550/arXiv.2505.11263},
archivePrefix = {arXiv},
       eprint = {2505.11263},
 primaryClass = {astro-ph.GA},
       adsurl = {https://ui.adsabs.harvard.edu/abs/2025arXiv250511263N}
}

@ARTICLE{2025NatAs...9.1134A,
       author = {Adamo+},
title = {},
      journal = {Nature Astronomy},
         year = 2025,
       volume = {9},
        pages = {1134-1147},
          doi = {10.1038/s41550-025-02624-5},
archivePrefix = {arXiv},
       eprint = {2405.21054},
 primaryClass = {astro-ph.GA},
       adsurl = {https://ui.adsabs.harvard.edu/abs/2025NatAs...9.1134A}
}

@ARTICLE{2023NatAs...7..731B,
       author = {Boylan-Kolchin+},
title = {},
      journal = {Nature Astro.},
         year = 2023,
       volume = {7},
        pages = {731-735},
          doi = {10.1038/s41550-023-01937-7},
archivePrefix = {arXiv},
       eprint = {2208.01611},
 primaryClass = {astro-ph.CO},
       adsurl = {https://ui.adsabs.harvard.edu/abs/2023NatAs...7..731B}
}

@ARTICLE{2003ApJ...587L..79F,
       author = {Franx+},
title = {},
      journal = {\apjl},
         year = 2003,
       volume = {587},
       number = {2},
        pages = {L79-L82},
          doi = {10.1086/375155},
archivePrefix = {arXiv},
       eprint = {astro-ph/0303163},
 primaryClass = {astro-ph},
       adsurl = {https://ui.adsabs.harvard.edu/abs/2003ApJ...587L..79F}
}

@ARTICLE{2020A&A...641A...6P,
       author = {Planck Coll.},
      journal = {\aap},
         year = 2020,
       volume = {641},
          eid = {A6},
        pages = {A6},
          doi = {10.1051/0004-6361/201833910},
archivePrefix = {arXiv},
       eprint = {1807.06209},
 primaryClass = {astro-ph.CO},
       adsurl = {https://ui.adsabs.harvard.edu/abs/2020A&A...641A...6P}
}

@ARTICLE{2005PhR...405..279B,
       author = {Bertone+},
      journal = {\physrep},
         year = 2005,
       volume = {405},
       number = {5-6},
        pages = {279-390},
          doi = {10.1016/j.physrep.2004.08.031},
archivePrefix = {arXiv},
       eprint = {hep-ph/0404175},
 primaryClass = {hep-ph},
       adsurl = {https://ui.adsabs.harvard.edu/abs/2005PhR...405..279B}
}

@ARTICLE{TSM22,
       author = {Treu+},
      journal = {\aapr},
         year = 2022,
       volume = {30},
       number = {1},
          eid = {8},
        pages = {8},
          doi = {10.1007/s00159-022-00145-y},
archivePrefix = {arXiv},
       eprint = {2210.15794},
 primaryClass = {astro-ph.CO},
       adsurl = {https://ui.adsabs.harvard.edu/abs/2022A&ARv..30....8T}
}

@ARTICLE{ZNT24,
       author = {Zelko+},
      journal = {\mnras},
         year = 2024,
       volume = {531},
       number = {1},
        pages = {885-897},
          doi = {10.1093/mnras/stae970},
archivePrefix = {arXiv},
       eprint = {2311.17140},
 primaryClass = {astro-ph.CO},
       adsurl = {https://ui.adsabs.harvard.edu/abs/2024MNRAS.531..885Z}
}

@ARTICLE{Divalentino2025,
       author = {Di Valentino+},
      journal = {Physics of the Dark Universe},
         year = 2025,
       volume = {49},
          eid = {101965},
        pages = {101965},
          doi = {10.1016/j.dark.2025.101965},
archivePrefix = {arXiv},
       eprint = {2504.01669},
 primaryClass = {astro-ph.CO},
       adsurl = {https://ui.adsabs.harvard.edu/abs/2025PDU....4901965D}
}

@ARTICLE{2022Natur.604..261F,
       author = {Fujimoto+},
      journal = {\nat},
         year = 2022,
       volume = {604},
       number = {7905},
        pages = {261-265},
          doi = {10.1038/s41586-022-04454-1},
archivePrefix = {arXiv},
       eprint = {2204.06393},
 primaryClass = {astro-ph.GA},
       adsurl = {https://ui.adsabs.harvard.edu/abs/2022Natur.604..261F}
}

@ARTICLE{2024Natur.627...59M,
       author = {Maiolino+},
      journal = {\nat},
         year = 2024,
       volume = {627},
       number = {8002},
        pages = {59-63},
          doi = {10.1038/s41586-024-07052-5},
archivePrefix = {arXiv},
       eprint = {2305.12492},
 primaryClass = {astro-ph.GA},
       adsurl = {https://ui.adsabs.harvard.edu/abs/2024Natur.627...59M}
}

@ARTICLE{2023ApJ...953L..29L,
       author = {Larson+},
      journal = {\apjl},
         year = 2023,
       volume = {953},
       number = {2},
          eid = {L29},
        pages = {L29},
          doi = {10.3847/2041-8213/ace619},
archivePrefix = {arXiv},
       eprint = {2303.08918},
 primaryClass = {astro-ph.GA},
       adsurl = {https://ui.adsabs.harvard.edu/abs/2023ApJ...953L..29L}
}

@ARTICLE{2025arXiv250400075S,
       author = {Shen+},
      journal = {arXiv:2504.00075},
         year = 2025,
          eid = {arXiv:2504.00075},
          doi = {10.48550/arXiv.2504.00075},
archivePrefix = {arXiv},
       eprint = {2504.00075},
 primaryClass = {astro-ph.GA},
       adsurl = {https://ui.adsabs.harvard.edu/abs/2025arXiv250400075S}
}

@ARTICLE{2024MNRAS.527.2835S,
       author = {Shen+},
      journal = {\mnras},
         year = 2024,
       volume = {527},
       number = {2},
        pages = {2835-2857},
          doi = {10.1093/mnras/stad3397},
archivePrefix = {arXiv},
       eprint = {2304.06742},
 primaryClass = {astro-ph.GA},
       adsurl = {https://ui.adsabs.harvard.edu/abs/2024MNRAS.527.2835S}
}

@ARTICLE{2024MNRAS.530.4868Y,
       author = {Yung+},
      journal = {\mnras},
         year = 2024,
       volume = {530},
       number = {4},
        pages = {4868-4886},
          doi = {10.1093/mnras/stae1188},
archivePrefix = {arXiv},
       eprint = {2309.14408},
 primaryClass = {astro-ph.CO},
       adsurl = {https://ui.adsabs.harvard.edu/abs/2024MNRAS.530.4868Y}
}

@ARTICLE{2024ApJ...976...40W,
       author = {Winch+},
      journal = {\apj},
         year = 2024,
       volume = {976},
       number = {1},
          eid = {40},
        pages = {40},
          doi = {10.3847/1538-4357/ad7a73},
archivePrefix = {arXiv},
       eprint = {2404.11071},
 primaryClass = {astro-ph.CO},
       adsurl = {https://ui.adsabs.harvard.edu/abs/2024ApJ...976...40W}
}

@ARTICLE{2025A&A...697A..65M,
       author = {Matteri+},
      journal = {\aap},
         year = 2025,
       volume = {697},
          eid = {A65},
        pages = {A65},
          doi = {10.1051/0004-6361/202553701},
archivePrefix = {arXiv},
       eprint = {2503.01968},
 primaryClass = {astro-ph.GA},
       adsurl = {https://ui.adsabs.harvard.edu/abs/2025A&A...697A..65M}
}

@ARTICLE{2023JCAP...10..072A,
       author = {Adil+},
      journal = {\jcap},
         year = 2023,
       volume = {2023},
       number = {10},
          eid = {072},
        pages = {072},
          doi = {10.1088/1475-7516/2023/10/072},
archivePrefix = {arXiv},
       eprint = {2307.12763},
 primaryClass = {astro-ph.CO},
       adsurl = {https://ui.adsabs.harvard.edu/abs/2023JCAP...10..072A}
}

@ARTICLE{2025ApJ...988..264M,
       author = {Manzoni+},
      journal = {\apj},
         year = 2025,
       volume = {988},
       number = {2},
          eid = {264},
        pages = {264},
          doi = {10.3847/1538-4357/ade700},
archivePrefix = {arXiv},
       eprint = {2502.04702},
 primaryClass = {astro-ph.GA},
       adsurl = {https://ui.adsabs.harvard.edu/abs/2025ApJ...988..264M}
}

\end{multicols}

\end{document}